\documentclass[aps,twocolumn,superscriptaddress,hyperref]{revtex4}
\usepackage{graphicx,psfrag,amsmath,times,hhline}
\begin{document}

\title{Lossless Compression and Complexity of Chaotic Sequences}
\author{Nithin Nagaraj}
\author{Mathew Shaji Kavalekalam}
\author{Arjun Venugopal T.}
\author{Nithin Krishnan}
\affiliation{Department of Electronics and Communications, Amrita
School of Engineering, Amrita Vishwa Vidyapeetham, Amritapuri
Campus, Clappana P.O., Kollam, Kerala - 690 525.}

\begin{abstract}
We investigate the complexity of short symbolic sequences of chaotic
dynamical systems by using lossless compression algorithms. In
particular, we study Non-Sequential Recursive Pair Substitution
(NSRPS), a lossless compression algorithm first proposed by W.
Ebeling {\it et al.} [Math. Biosc. 52, 1980] and
Jim\'{e}nez-Monta\~{n}o {\it et al.} [arXiv:cond-mat/0204134, 2002])
which was subsequently shown to be optimal. NSPRS has also been used
to estimate Entropy of written English (P. Grassberger
[arXiv:physics/0207023, 2002]).  We propose a new measure of
complexity - defined as the number of iterations of NSRPS required
to transform the input sequence into a constant sequence. We test
this measure on symbolic sequences of the Logistic map for various
values of the bifurcation parameter. The proposed measure of
complexity is easy to compute and is observed to be highly
correlated with the Lyapunov exponent of the original non-linear
time series, even for very short symbolic sequences (as short as 50
samples). Finally, we construct symbolic sequences from the
Skew-Tent map which are incompressible by popular compression
algorithms like WinZip, WinRAR and 7-Zip, but compressible by NSRPS.
\end{abstract}

\maketitle

\section{Introduction}
\label{secIntro} %
Measuring complexity of experimental time series is one of the
important goals of mathematical modeling of natural phenomena. A
measure of complexity gives an insight in to the phenomenon being
studied. For example, in a study of population dynamics of the
fruit-fly, a measure of complexity of the time series (population
size of generations) will throw light on the persistence and stability of the population. If
the complexity is low, then it is possible that the population is exhibiting a periodic behavior,
i.e. fluctuating between a high population size and a low one
alternately. Complexity also plays a very important role in
determining whether a sequence is random or not in Cryptography
applications.

Different measures of complexity such as Lyapunov Exponent,
Kolmogorov complexity, Algorithmic complexity etc. are proposed in
the literature~\cite{Measures}. While complexity has several facets,
Shannon entropy~\cite{Shannon48} is one of the reliable indicators
of `compressibility' which can serve as a measure of complexity. It
is given by the following expression:
\begin{equation}
H(X) = -\sum_{i=1}^{M} p_i \log_2(p_i)~~~~\text{bits/symbol},
\end{equation}
where $X$ is the symbolic sequence with $M$ symbols and $p_i$ is the
probability of the $i$-th symbol for a block-size of one. Block-size refers to the number of
input symbols taken together to compute the probability mass
function.

Shannon entropy plays an important role in lossless data storage
and communications. Shannon's Noiseless Source Coding
theorem~\cite{Shannon48} provides an upper limit on the compression
ratio achievable by lossless compression algorithms. This limit is
given by the Shannon entropy. Numerous algorithms have been
designed with the aim of achieving this limit. Huffman coding,
Shannon-Fano coding, Arithmetic coding, Lempel-Ziv coding are a few
examples of lossless compression algorithms which achieve the
Shannon entropy limit for stochastic i.i.d sources (independent and
identically distributed)~\cite{Sayood, Cover}. However, practical
estimation of entropy of sources is non-trivial since most sources
are not i.i.d but contain correlations (short or long-range). As a
simple example, in the English language, the probability of the
occurrence of the letter `$u$' after the letter `$q$' has occurred,
is nearly one.

In this paper, we are interested in measuring complexity of short
symbolic sequences which are obtained from time series generated by
chaotic non-linear dynamical systems (we have used the Logistic map
in our study and we expect the results to hold for other systems as
well).

This paper is organized as follows. In the next section, we
highlight the challenges in measuring an estimate of Shannon
entropy for short sequences. In section III, we introduce NSRPS and
propose a new measure of complexity based on this algorithm.
Subsequently, in section IV, we test the new measure on several
(short) sequences from the Logistic map and compare the complexity
with a uniformly distributed random sequence~\footnote{Random
sequences are characterized by maximum complexity as well as maximum
Shannon entropy.}. The complexity measure based on NSRPS is compared
with Lyapunov Exponent.  In section V, we construct chaotic
sequences which are incompressible by popular lossless compression
algorithms, but which can be compressed by NSRPS. We conclude in
section VI indicating directions for future work.

\section{Drawbacks of Shannon Entropy as a Complexity Measure}
\label{secLossless} %
Shannon entropy can serve as a good indicator for complexity, but
estimation of entropy is not a trivial task. Determining Shannon
entropy of experimental time series is particularly challenging
owing to the following reasons:

\begin{enumerate}
\item Analytical determination of the entropy is not easy even for a
simple model of the experimental time series.

\item The time series typically consists of real numbers. In order to calculate the entropy, it
has to be converted into a symbolic sequence. The choice of the
partition has a very important role to play in the estimation of the
entropy. Ebeling {\it et al.}~\cite{Ebeling} shows that depending on
the choice of the partition, the results can vary widely.

\item Noise is inevitable in any experiment. Noise has the tendency
to increase entropy.

\item Length of the time series is another important factor in the
accurate determination of entropy. Shannon entropy requires the
estimation of the probability mass function, which is difficult
to accurately estimate with a short time series. Biological time
series such as population sizes are typically of very small
lengths, around 50-100 samples (since actual experiments are time
consuming). Entropy estimation methods in literature require 1000 to
10000 samples~\cite{Ebeling}.
\end{enumerate}

In order to overcome these drawbacks, researchers have used lossless
compression algorithms in order to estimate complexity or
entropy~\cite{Puglisi, Grassberger, Entropy1}. Lempel-Ziv and its
popular variations are extensively used by several researchers to
determine complexity of time series (\cite{Puglisi} and references
therein). As we shall demonstrate in section V, this is not always
reliable for short sequences.

\subsection{Effect of length of time series on Entropy estimation}
\begin{figure}[!h]
\begin{center}
\centering
\includegraphics[scale=0.45]{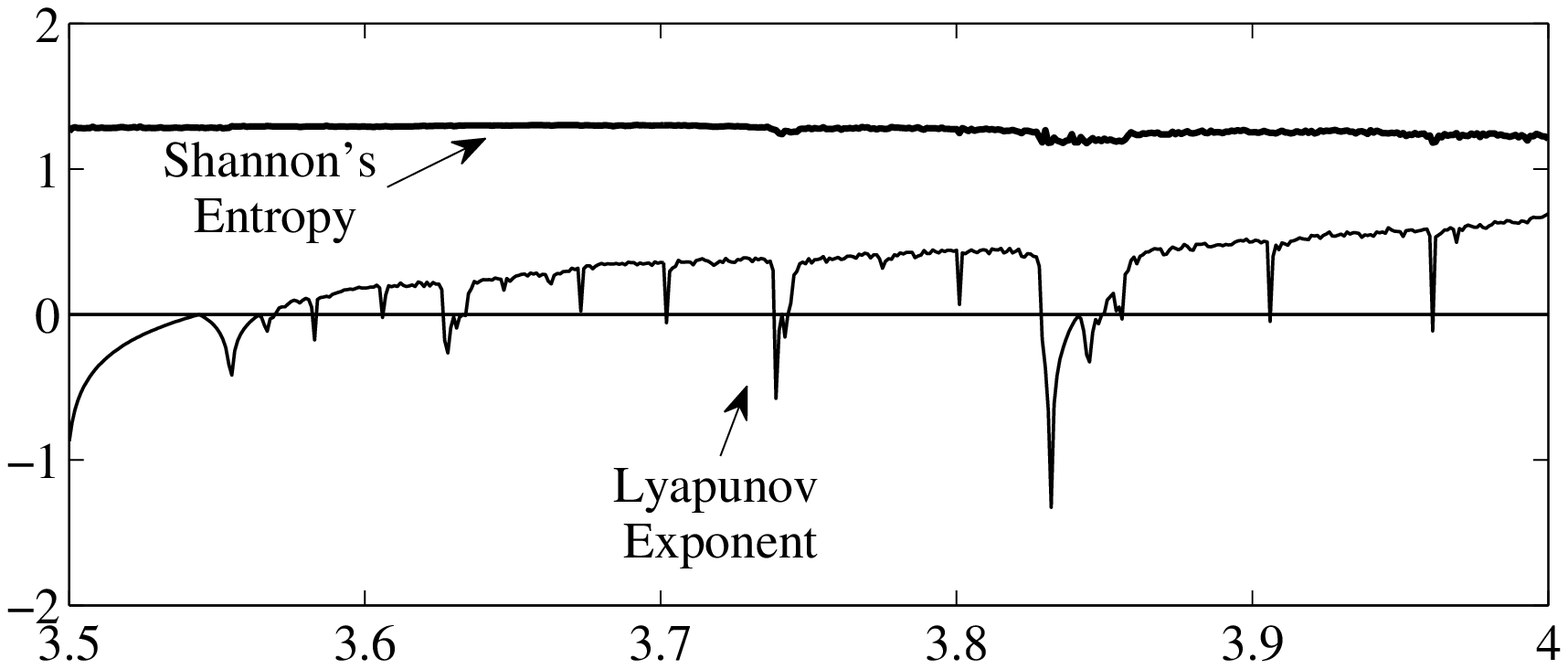}
\caption[Ent and LE]{Numerically computed Shannon entropy vs.
Lyapunov Exponent $\lambda$ as the bifurcation parameter is varied.
8 bins~\footnote{The 8 bins are uniformly spaced between 0 to 1 and
the input time series is converted into a symbolic sequence
consisting of 8 symbols corresponding to these bins.} were used for
the numerical computation of Shannon entropy using equation (1)
and data-length $L=200$. For computation of $\lambda$, equation (3)
was used. The two graphs are poorly correlated as indicated by a
correlation coefficient of -0.2682.} \label{figure:entLE}
\end{center}
\end{figure}

Fig.~\ref{figure:entLE} shows the effect of length of the time series on numerical
computation of the Shannon entropy. For a data-length $L=200$, as
the bifurcation parameter of the Logistic map is varied from $3.5$
to $4.0$, we observe that the numerically estimated Shannon
entropy (equation (1)) is poorly correlated with Lyapunov exponent
with a correlation coefficient of $-0.2682$. When the data-length is
increased to $L=5000$, Shannon entropy comes close to the Lyapunov
exponent with a correlation coefficient of 0.8934.
Ebeling~\cite{Ebeling} demonstrates that for the Logistic map,
Shannon entropy comes very close to the Lyapunov exponent as the
block-size increases to 10 and for large data-lengths
($L \ge 1000$).

\section{NSRPS-based measure of complexity}
\label{secMeasure} %
In this section, we propose a new measure of complexity based on a
lossless compression algorithm called Non-sequential Recursive Pair
Substitution (NSRPS). NSRPS was first proposed by Ebeling {\it et
al.}~\cite{nsrps1} and later improved by Jim\'{e}nez-Monta\~{n}o
{\it et al.}~\cite{nsrps2}. It was subsequently shown to be
optimal~\cite{nsrps3}. NSPRS has also been used to estimate Entropy
of written English~\cite{Grassberger}. The algorithm is briefly
described as follows.

Let the original sequence be called $X$. At the first iteration,
find which pair of symbols have maximum number of occurrences and
replace all its non-overlapping occurrences with a new symbol. For
example, the input sequence `$01101011$' is transformed into
`$21221$' since the pair `$01$' has maximum number of occurrences
compared to other pairs (`$00$', `$10$' and `$11$'). In the second
iteration, `$21221$' is transformed to `$323$' since `$21$' has
maximum frequency. The algorithm proceeds in this fashion until the
length of the string is 1 (at which stage there is no pair to
substitute). In this example, in the third iteration, `$323$' is
transformed into `$43$' and in the fourth iteration it is
transformed into `$5$' and the algorithm stops.

The following observations can be made about the algorithm:
\begin{enumerate}
\item The algorithm always terminates for finite length sequences.

\item After each iteration, the length of the sequence reduces. The
number of distinct symbols may or may not increase (if the input
sequence is `$0000$', then it is transformed to `$22$' and then to
`$3$').

\item The quantity `$entropy \times length$' may increase or
decrease across the iterations.

\item Ultimately, the quantity `$entropy \times length$' has to go
to zero since the length eventually reaches 1 at which point the
entropy is 0 (since there is now only one symbol, it occurs with
probability 1). A faster way for this quantity to go to zero is when
the sequence gets transformed to a constant sequence (which has only
one distinct symbol and hence zero entropy). Let $N$ be the number
of iterations required for the quantity `$entropy \times length$' to
reach zero. $N$ is always a positive integer. The minimum value of
$N$ is zero (for the constant sequence) and maximum is $L-1$ where $L$ is the length of the sequence (for a sequence either with distinct symbols are with all pairs being distinct).

\item The algorithm as described above is not reversible, i.e. the original symbolic
sequence can't be restored by the sequence at subsequent iterations.
In order to make the algorithm reversible, we have to maintain a
record of the specific pair of symbols which was substituted at each
iteration. The bits required to store this overhead information
compensates for the reduction in the number of bits needed to store
the transformed sequence. For achieving the best lossless
compression ratio, we stop at the iteration number at which the
total number of bits required to store the transformed sequence and
the overhead is a minimum (and hopefully lesser than the size of the
original sequence).
\end{enumerate}

\subsection{Definition of the new complexity measure}
The number of iterations $N$ for the quantity `$entropy \times
length$' to approach zero by the NSRPS algorithm (as described
above) is defined as our new complexity measure. $N$ is an integer in the range $[0, L-1]$.

Jim\'{e}nez-Monta\~{n}o~\cite{nsrps2} actually tracks the quantity
`$entropy \times length$' across the iterations of NSRPS. While this
is important, our motivation to use $N$ as a complexity measure is
the following. $N$ actually represents the {\it effort} required by
NSRPS algorithm to transform the input sequence into a constant
sequence (having only one distinct symbol and hence zero entropy). A
sequence which is highly redundant would naturally have a lower
value of $N$. As an example, the sequences $A=01010101$ and
$B=01001110$ have the same length ($8$) and the same entropy of 1
bits/symbol (block-size=1). However, sequence $A$ requires only
$N=1$ iteration for the quantity $entropy \times length$ to reach
zero, whereas $B$ requires $N=6$ iterations~\footnote{$A=01010101$
$\mapsto$ $2222$ $\implies$ $N=1$. Now $B=01001110$ $\mapsto$
$202110$ $\mapsto$ $32110$ $\mapsto$ $4110$ $\mapsto$ $510$
$\mapsto$ $60$ $\mapsto$ $7$ $\implies$ $N=6$.}. Clearly, $B$ is
more {\it complex} than $A$ ($A$ is periodic, $B$ has no
obvious pattern).

\section{Results and Discussion}
\label{secResults} %
In this section, we shall evaluate the usefulness of the new
complexity measure based on NSRPS described in the previous section.
To this end, we consider sequences arising from the Logistic map for
various values of the bifurcation parameter `$a$'. We know that the
complexity of the time series increases with `$a$', with occasional
dips owing to the presence of {\it windows} (attracting periodic
orbits).

\begin{figure}[!h]
\centering
\includegraphics[scale=0.45]{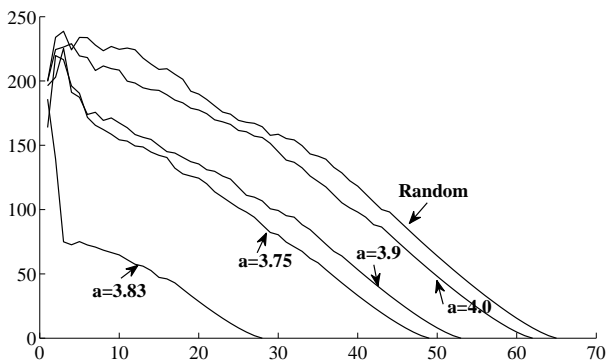}
\caption[NSRPS]{Complexity of different sequences ($L=200$):
$entropy \times length$ vs. number of iterations.}
\label{figure:nsrpsComp}
\end{figure}

\begin{figure}[!h]
\centering
\includegraphics[scale=0.45]{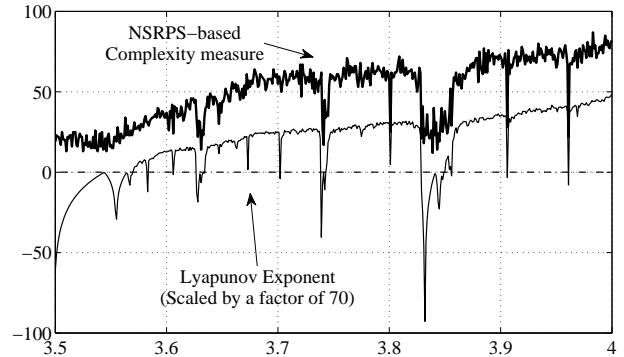}
\caption[NSRPS and LE]{NSRPS based complexity measure $N(a)$ vs.
Lyapunov Exponent $\lambda(a)$ as the bifurcation parameter `$a$' is
varied between 3.5 to 4.0. We have used 8 bins for deriving the
symbolic sequence from the time series. The data-length is $L=200$.
For computation of $\lambda(a)$ we have used equation (3).
$\lambda(a)$ was scaled by a factor of 70. The two graphs are highly
correlated as indicated by a correlation coefficient of 0.8832.
Compare this with Fig.~\ref{figure:entLE}.} \label{figure:nsrpsLE}
\end{figure}

In Fig.~\ref{figure:nsrpsComp}, the quantity $entropy \times
length$ is plotted along the Y-axis and iteration number along the
X-axis. The length of all sequences is $L=200$. The new complexity
measure $N$ is the iteration number when the graph hits X-axis. As
it can be seen, different sequences have different values of $N$. As
expected, the sequence with the highest complexity is the
independent and uniformly distributed random sequence ($rand()$ in
MATLAB). The order of complexity (from higher to lower) is $random
\succ a=4.0 \succ 3.9 \succ 3.75 \succ 3.83$. There is an attracting
periodic orbit ({\it window}) at $a=3.83$ and this explains the
lower value of $N$.

Table~1 shows the effect of data-length and number of bins on the
new measure $N$ for the Logistic map. As we vary the bifurcation parameter `$a$' between
3.5 to 4.0, we find that even for $L=50$, the correlation
coefficient (CC) of $N(a)$ with the Lyapunov Exponent $\lambda(a)$
is quite good. The entropy $H$ (calculated using equation (1)) is
very poorly correlated with $\lambda$. For 2 bins, even at $L=5000$,
we found the CC of $H(a)$ and $\lambda(a)$ to be 0.3565. Compare
this with Table~1: for $L=50$ and 2 bins, the CC is already 0.6651.
This shows that the new measure is quite good for very short
symbolic sequences. Figure~\ref{figure:nsrpsLE} shows the graphs of
$N(a)$ and $\lambda(a)$ (scaled by a factor of 70 for better
visibility and ease of comparison).
\begin{table}[!h]
\begin{center}
\caption{Effect of data-length and number of bins on the new measure
$N$ in comparison with Lyapunov Exponent ($\lambda$). CC stands for
correlation coefficient between $N(a)$ and $\lambda(a)$ as the
bifurcation parameter `$a$' of the Logistic map is varied from 3.5
to 4.0.} \label{tableComp} \vskip 10pt
\begin{tabular}{p{1cm} p{1.5cm} p{1cm}}
\hhline{===}
    L & \# of Bins & CC  \\
  \hline
      & 2 & 0.6651  \\
  50  & 4 & 0.6654  \\
      & 8 & 0.7324  \\
  \hline
       & 2 & 0.8352  \\
  100  & 4 & 0.8149  \\
       & 8 & 0.8172  \\
  \hline
      & 2 & 0.8870  \\
  200 & 4 & 0.8648  \\
      & 8 & 0.8832  \\
\hhline{===}
\end{tabular}
\end{center}
\end{table}

\noindent Lyapunov Exponent $\lambda$ is given by the equation:
\begin{equation}
\lambda = \lim_{n \rightarrow \infty} \frac{1}{n}
\sum_{i=1}^{n}\ln(|f'(x_i)|).
\end{equation}
For the Logistic map, we have used the following equation to
estimate $\lambda$:
\begin{equation}
\lambda(a) = \frac{1}{1000} \sum_{i=1}^{1000}\ln(|a(1-2x_i)|),
\end{equation}
where `$a$' is the bifurcation parameter (3.5 to 4.0) and $x_1$ is a
randomly chosen initial condition in the interval (0,1).

 The number of bins determines the number of symbols for the
initial sequence. As $L$ and number of bins increase, the CC gets
better and better.

\section{Chaotic Sequences from Skew-Tent Map}
\label{secSkewTent} %
Complexity measures based on lossless data compression are not
always accurate, especially for short data lengths, as we shall
demonstrate. Consider the Skew-tent map~\cite{skewtent}:
\begin{eqnarray*}\label{Eqskew}
    x_{i+1} & = & \frac{x_i}{a}~~~~~~~ \text{if $0 \leq x_i \leq a$},\\
    & = & \frac{1-x_i}{1-a}~~~ \text{if $a < x_i \leq 1$}.
\end{eqnarray*}
Here `$a$' can be any value in the interval [0.5,1). For $a=0.5$, we
have the well-known Tent map.

Using the value $a=0.65$, data length $L=1024$ and using a random
initial condition, we first obtain a chaotic time series. From this,
we find the symbolic sequence with 2 bins. The first bin is $[0,
0.5)$ corresponding to symbol `0' and the second bin $[0.5, 1)$
corresponding to symbol `1'. The symbols `0' and `1' are equally
likely since the invariant distribution for the Skew-tent map is
uniform~\cite{skewtent}. This implies that the Shannon entropy is
1 bits/symbol.

For compression using NSRPS, the overhead information was taken in to account. Table~\ref{tableCompression} shows the efficacy of NSRPS
for compressing such chaotic sequences of short length while other
popular compression algorithms expand (all these use some variation of Lempel-Ziv compression algorithm). This behaviour was observed for values of $a$
between 0.5 and 0.7. Rigorous investigation of these interesting
sequences needs to be performed.
\begin{table}[!h]
\begin{center}
\caption{Chaotic sequences from the Skew-tent map subjected to
lossless compression algorithms. All numbers are in bits. As it can
be seen, only NSRPS manages to compress the sequence.}
\label{tableCompression} \vskip 5pt
\begin{tabular}{|c|c|c|c|c|}
\hline
    Input size & WinZip & WinRAR & 7-Zip & NSRPS \\
    \hline
     1024      & 1616   & 1376   & 1816  & 912 \\
  \hline
\end{tabular}
\end{center}
\end{table}
\section{Conclusions and Future Work}
\label{secConclusions} %
The new measure is able to correctly characterize the
complexity of chaotic sequences as demonstrated for the Logistic map
(for different values of the bifurcation parameter) and a uniformly
distributed random sequence. This new measure is highly correlated
with the Lyapunov exponent even for very small data-lengths, as low
as $L=50$.


Future work would be to investigate the effect of various kinds of
noise (corrupting the time series) on the complexity measure $N$. We
have reasons to believe that $N$ would be robust to noise to some
extent since we are working on the symbolic sequence. The new
measure needs to be further tested for various dynamical systems
(maps and flows) and stochastic time series of different
distributions, and to non-uniform bin structures. The data
compression aspect of NSRPS needs to the thoroughly investigated,
especially for compressing chaotic sequences which are otherwise
incompressible by standard techniques.$\vspace{0.1in}$

\noindent {\bf{Acknowledgments:}} The authors express their
heart-felt gratitude to Mata Amritanandamayi Devi (affectionately
known as `Amma' which means `Mother') for her constant support in
material and spiritual matters. NN thanks Sutirth Dey (IISER, Pune) for useful discussions and Department of Biotechnology, Govt. of India for funding through the RGYI scheme.

\end{document}